\documentclass[12pt]{article}

\newcommand{\sect}[1]{\setcounter{equation}{0}\section{#1}}

\global\arraycolsep=1pt
\def\fnote#1#2{\begingroup\def\thefootnote{#1}\footnote{#2}\addtocounter
{footnote}{-1}\endgroup}

\usepackage{amsbsy,amssymb,latexsym,epsfig}

\newcommand{\beq}{\begin{equation}}
\newcommand{\eeq}{\end{equation}}
\newcommand{\beqa}{\begin{eqnarray}}
\newcommand{\eeqa}{\end{eqnarray}}
\newcommand{\bR}{{\mathbb R}}
\newcommand{\bRP}{{\mathbb{RP}}}
\newcommand{\bZ}{{\mathbb Z}}

\newcommand{\CP}{{\mathbb{CP}}}
\newcommand{\CN}{{\mathcal N}}

\def\fnote#1#2{\begingroup\def\thefootnote{#1}\footnote{#2}\addtocounter
{footnote}{-1}\endgroup}

\usepackage{amsbsy,amssymb,latexsym}

\begin{document}
\begin{flushright}
OCU-PHYS 222 \\
hep-th/0411165

\end{flushright}
\vspace{12mm}

\begin{center}
{\bf\Large
Seven-dimensional Einstein Manifolds\\
from
Tod-Hitchin Geometry 
}

\vspace{15mm}
Makoto Sakaguchi\fnote{$\star$}{
\texttt{msakaguc@sci.osaka-cu.ac.jp}
}
and
Yukinori Yasui\fnote{$\ast$}{
\texttt{yasui@sci.osaka-cu.ac.jp}
}
\vspace{10mm}

\textit{
${{}^\star}$
Osaka City University
Advanced Mathematical Institute (OCAMI)
}
\vspace{2mm}

${}^\ast$
\textit{
Department of Mathematics and Physics,
Graduate School of Science,\\
Osaka City University
}

\vspace{3mm}

\textit{
Sumiyoshi,
Osaka 558-8585, JAPAN}
\end{center}
\vspace{8mm}

\begin{abstract}
We construct infinitely many seven-dimensional Einstein metrics
of weak holonomy $G_2$. These metrics are defined on principal
SO(3) bundles over four-dimensional Bianchi IX orbifolds with
the Tod-Hitchin metrics. The Tod-Hitchin metric has an orbifold
singularity parameterized by an integer,
and is shown to be similar near the singularity
to the Taub-NUT de Sitter metric with a special charge.
We show, however, 
that the seven-dimensional metrics on the total space
 are actually smooth.
The geodesics on the weak $G_2$ manifolds are discussed.
It is shown that the geodesic equation is equivalent to the
Hamiltonian equation
of an interacting rigid body system.
We also discuss M-theory on
the product space of AdS$_4$ and
the seven-dimensional manifolds,
and the dual gauge theories in three-dimensions.
\end{abstract}
\newpage

%%%%%%%%%%%%%%%%%%%%%%%%%%%%%%%%%%%%%%%%%%%%%%%%%%%%%%%%%%%%%%%%%%%%
\sect{Introduction}

M-theory compactifications on special holonomy manifolds
have attracted much attention,
because they preserve some supersymmetry
and allow to examine
dynamical aspects of a large class of supersymmetric gauge theories
\cite{Acharya:2004qe}.
For example,
it is known that there are eight-dimensional Ricci flat
manifolds with holonomy
Sp(2), SU(4) and Spin(7) except for the trivial one,
and M-theory compactifications on them correspond to
three-dimensional gauge theories
with $\CN=3$, $2$ and $1$ supersymmetry, 
respectively.
For a non-compact eight-dimensional
 special holonomy manifold,
M-theory on it is interpreted as
a worldvolume theory on
an M2-brane with
a special holonomy manifold as
the transverse space.
This is closely related to the supersymmetric 
M-theory solution AdS$_4\times M$
with compact 
seven-dimensional Einstein manifold $M$.
For 
weak $G_2$ manifolds $M$,
namely,
3-Sasakian, Sasaki-Einstein and proper weak $G_2$ manifolds,
the M-theory solutions 
AdS$_4\times M$
are AdS/CFT dual to $\CN=3$, $2$ and $1$ superconformal field theories
on the boundary of AdS$_4$
\cite{Maldacena}\cite{Gubser:1998bc}\cite{Witten:1998qj}\cite{Aharony:1999ti}.
The brane solution
naturally interpolates between AdS$_4\times M$ in the near horizon limit
and $\bR^{1,2} \times C(M)$,
 where $C(M)$ is the cone
over $M$ with the special holonomy SP(2), SU(4) or Spin(7),
and the gauge theories on the both sides are related by the RG-flow
\cite{AFHS}.

In this paper,
we construct infinitely many seven-dimensional Einstein metrics admitting
3-Sasakian and proper weak $G_2$ structures
\fnote{$\flat$}{
Recently, infinitely many Sasaki-Einstein metrics are
constructed in \cite{Gauntlett:2004hh}\cite{Boyer-Galicki}
}.
These metrics are defined on compact manifolds $M_k$
parameterized by an integer $k\ge 3$; principal
SO(3) bundles over four-dimensional Bianchi IX orbifolds with
the Tod-Hitchin metrics
\cite{Tod}\cite{Hitchin:twistor}\cite{Hitchin:a new family}. 
The Tod-Hitchin metric has an orbifold
singularity parameterized by the integer $k$. 
However, 
the singularity is resolved by adding the fiber SO(3),
and so the total spaces $M_k$ become smooth manifolds.
Our compact manifolds 
contain manifolds $S^7$, $N^{0,1,0}$ and the squashed S$^7$
as special homogeneous cases for $k=3,4$
\cite{Duff:1986hr}.
For generic $k$, the metrics on $M_k$ are inhomogeneous
and admit SO(3)$\times$SO(3) isometry.
This implies that
the dual gauge theories in three-dimensions
are $\CN=3$ supersymmetric with SO(3) flavor for 3-Sasakian manifolds
$M_k$,
and $\CN=1$ supersymmetric with SO(3)$\times$SO(3) flavor for proper
weak $G_2$ manifolds $M_k$.
We examine the geodesics on $M_k$ using a Hamiltonian formulation
on the cotangent bundle $T^{*}M_k$.
The geodesic equation is equivalent to the Hamiltonian equation
of an interacting rigid body system. 
We find some special solutions,
which may be useful to consider the Penrose limit of our metrics.

This paper is organized as follows.
In section 2,
we introduce the Tod-Hitchin geometry,
and explain the relation to the Atiyah-Hitchin manifold \cite{Atiyah:1988jp}.
We show that the Tod-Hitchin geometry
is well approximated by the Taub-NUT de-Sitter geometry with a special charge.
In section 3,
we construct infinitely many seven-dimensional Einstein metrics
of weak holonomy $G_2$ on compact manifolds.
We also discuss the geodesics on the weak $G_2$ manifolds,
in section 4.
In the last section, 
we comment on the M-theory solutions AdS$_4\times M_k$
and the dual gauge theories in three-dimensions.
In appendix A, we present the anti-self-dual condition for the
Bianchi IX Einstein metric.
We summarize the relation between the Tod-Hitchin metric
and the Painlev\'e VI solution in appendix B.
In appendix C, the $G_2$ structure of the metric is given.

%%%%%%%%%%%%%%%%%%%%%%%%%%%%%%%%%%%%%%%%%%%%%%%%%%%%%%%%%%%%%%%%%%%%

\sect{ASD Einstein metrics on four-dimensional Bianchi IX manifold}

In this section, we consider Bianchi IX Einstein metrics with positive cosmological constant.
By using the SO(3) left-invariant one-forms $\sigma_i$ $(i=1,2,3)$,
the metric can be written in the form:
\begin{eqnarray}
g=dt^2
+a^2(t)\sigma_1^2
+b^2(t)\sigma_2^2
+c^2(t)\sigma_3^2~.
\label{Bianchi IX}
\end{eqnarray}
In the biaxial case, the general solution
to the Einstein equation $Ric(g)=\Lambda g$
has three parameters,
the mass $m$, the NUT charge $n$ and the cosmological constant $\Lambda$;
\begin{eqnarray}
g_{\{m,n,\Lambda\}}=\frac{r^2-n^2}{\Delta(r)}dr^2
+\frac{4n^2\Delta(r)}{r^2-n^2}\sigma_1^2
+(r^2-n^2)(\sigma_2^2+\sigma_3^2),
\label{metric:TN}
\end{eqnarray}
where
\begin{eqnarray}
\Delta(r)=
r^2
-2mr
+n^2
+\Lambda\left(n^4+2n^2r^2-\frac{1}{3}r^4\right).
\end{eqnarray}

The anti-self-dual (ASD) condition for the Weyl curvature
determines $m$ in terms of
$n$ and $\Lambda$ as
\begin{eqnarray}
m=-n\left(1+\frac{4}{3}\Lambda n^2\right),
\end{eqnarray}
in which case 
\begin{eqnarray}
\Delta(r)=\frac{\Lambda}{3}(r+n)^2(r_+-r)(r-r_-)~,~~~
r_\pm=n\pm \sqrt{4n^2+\frac{3}{\Lambda}}~.
\end{eqnarray}
Then the metric (\ref{metric:TN})
 becomes the ASD Taub-NUT de-Sitter metric
\cite{CGLP}\cite{Gibbons:1978zy}
 given by
\begin{eqnarray}
g_{\{n,\Lambda\}}=\frac{dr^2}{F(r)}
+4n^2 F(r)\sigma_1^2
+(r^2-n^2)(\sigma_2^2+\sigma_3^2)~,
\label{Taub-NUT-de-Sitter}
\end{eqnarray}
where 
\begin{eqnarray}
F(r)=\frac{\Lambda}{3}\left(\frac{r+n}{r-n}\right)
(r_+-r)(r-r_-)~.
\end{eqnarray}
For $\Lambda=0$, the metric reduces to the ASD
Taub-NUT metric \cite{Hawking:1976jb},
\begin{eqnarray}
g_{\{n, 0\}}=\left(\frac{r-n}{r+n}\right)dr^2
+4n^2\left( \frac{r+n}{r-n} \right)\sigma_1^2
+(r^2-n^2)(\sigma_2^2+\sigma_3^2)~.
\label{TN}
\end{eqnarray}

We shall now restrict our attention to the
metric (\ref{Taub-NUT-de-Sitter}) with the
 special 
NUT charge
\begin{eqnarray}
n=\sqrt{\frac{3}{\Lambda (k^2-4)}}~,
\label{n}
\end{eqnarray}
which is
a family of
ASD
Einstein metrics $g_k\equiv g_{\{n=\sqrt{{3}/{\Lambda (k^2-4)}},\, \Lambda\}}$
parameterized by the integer $k\ge 3$.
Each metric $g_k$ 
has the following properties (see Figure 1):
\begin{description}
  \item[(a)] When the coordinate $r$ is taken to lie in the interval 
$n\le r \le r_+$, the metric has singularities at the boundaries;
There is an orbifold singularity at $r=r_+$, while
a curvature singularity at
another boundary $r=n$.
  \item[(b)] The metric gives an approximation to the
Tod-Hitchin metric.
  \item[(c)] As $k\to\infty$ and $\Lambda\to 0$
keeping $\Lambda k^2=3$,
 the metric converges to the ASD Taub-NUT
metric (\ref{TN}) with a negative mass parameter ($n=1$)
 which gives the asymptotic form of the
Atiyah-Hitchin hyperk\"ahler metric.
\end{description}
\begin{figure}
   \parbox{\textwidth}{
   \begin{center}
      \psfig{file=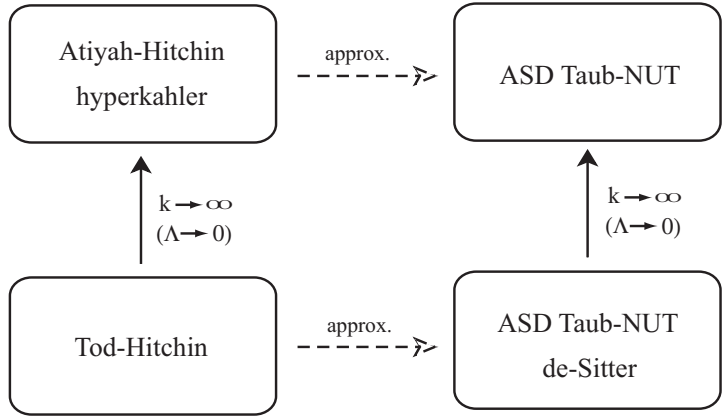,height=55mm}
   \end{center}
   \caption{The relation among metrics}
\label{fig:fig.1.eps}
}
\end{figure}

In the following, we will explain these points in some detail.
For this purpose,
we start with an explanation of
some relevant aspects of the Tod-Hitchin metrics.
Tod and Hitchin constructed a family of ASD Einstein metrics 
(Tod-Hitchin metrics)
on the Bianchi IX orbifold,
parameterized by an integer $k\ge  3$
\cite{Tod}\cite{Hitchin:twistor}\cite{Hitchin:a new family}.
These solutions are written in the triaxial form and have a compactification
as metrics with orbifold singularities.
These may be thought of as a resolution of the curvature singularity
in the ASD Taub-NUT de-Sitter metric $g_k$.
Each Tod-Hitchin metric $g^{\rm TH}_k$ is
 given by a solution to the Painlev\'e VI
equation (see appendix B).
For lower $k$ the metric takes the form
\cite{Hitchin:a new family}\cite{CGLP}:
\begin{itemize}
  \item $k=3$
\begin{eqnarray}
g^{\rm TH}_3=dt^2
+4\sin^2 t ~ \sigma_1^2
+4\sin^2(\frac{2}{3}\pi-t)~\sigma_2^2
+4\sin^2(t+\frac{2}{3}\pi)~\sigma_3^2 ~, \label{exact1}
\end{eqnarray}
which gives the standard metric on S$^4$ written in the
triaxial form.
  \item $k=4$
\begin{eqnarray}
g_4^{\rm TH}=dt^2
+\sin^2t~\sigma_1^2
+\cos^2 t~\sigma_2^2
+\cos^2 2 t~\sigma_3^2~,
\end{eqnarray}
which gives the Fubini-Study metric on $\CP^2$.
  \item $k=6,~8$~\\
The metric can be written as
\begin{eqnarray}
g^{\rm TH}_{k}=h(r)~ dr^2
+a^2(r)~\sigma_1^2
+b^2(r)~\sigma_2^2
+c^2(r)~\sigma_3^2~,
\end{eqnarray}
where the components are given
for $k=6$
\begin{eqnarray}
&&h^2 = \frac{3(1+r+r^2)}{r\, (r+2)^2\, (2r+1)^2}\,,\qquad
a^2 = \frac{3 (1+r+r^2) }{(r+2)\, (2r+1)^2}\,,\nonumber\\
&&b^2 =\frac{3r\, (1+r+r^2)}{(r+2)^2\, (2r+1)}\,,\qquad
c^2 =\frac{3(r^2-1)^2}{(1+r+r^2)\, (r+2)\, (2r+1)}~,
\end{eqnarray}
and for $k=8$
\begin{eqnarray}
h^2 &=& \frac{4(1+r)(3-2r+r^2)(1-2r+3r^2)(1+2r+3r^2)}{
   (1-r)\, r\, (1+r^2)(1+2r-r^2)^2\, (3+2r+r^2)^2}\,,\nonumber\\
a^2 &=& \frac{4(1-r)(1+r)^3\, (3-2r+r^2)(1-2r+3r^2)}{
         (1+2r-r^2)(3+2r+r^2)^2\, (1+2r+3r^2)}\,,\nonumber\\
b^2 &=& \frac{16r\, (1-2r+3r^2)(1+2r+3r^2)}{(1+2r-r^2)(3-2r+r^2)
(3+2r+r^2)^2}\,,\nonumber\\
c^2 &=& \frac{4\,(1+r^2) (3-2r+r^2)(1-2 r-r^2)^2(1+2r+3r^2)}{(1+2 r-r^2)^2(3+2r+r^2)^2(1-2r+3 r^2)}~.
\label{exact2}
\end{eqnarray}
\end{itemize}
Among the Tod-Hitchin metrics,
those with $k=3$ and $4$
are exceptional, i.e. there is no singularity.
The solutions with higher $k$
 are determined by the non-trivial solutions
to the Painlev\'e equation,
and in the limit $k\to \infty$ together with
a  suitable scaling of $\Lambda$
the solution approaches the Atiyah-Hitchin metric.
In the paper \cite{Hitchin:a new family},
 Hitchin found a systematic algebraic
way of finding solutions of the Painlev\'e equation.
However, it is not easy to write down these solutions explicitly.
To examine such a solution,
we consider the local metric near the boundary by using expansions
of
the solution (\ref{Bianchi IX}) to the
 Einstein equation.

To begin with, we discuss boundary conditions.
Let us impose a compact condition for the Bianchi IX
manifold 
$\simeq I\times $SO(3),
where $I$ is the closed interval $[t_1,t_2]\subset \bR$.
Furthermore
 we require that singularities at the boundaries,
 $t_1$ and $t_2$,
are described by bolts or nuts so that there are three types,
nut--nut, bolt--nut and bolt--bolt.
The Tod-Hitchin metric belongs to bolt--bolt type:
near $t=t_1$, the metric is written as
\begin{eqnarray}
g^{\rm TH}_k\sim dt^2
+\frac{4t^2}{(k-2)^2}\, \sigma_1^2
+L^2\, (\sigma_2^2+\sigma_3^2)~.
\label{t_1}
\end{eqnarray}
On the other hand, near $t=t_2$
\begin{eqnarray}
g^{\rm TH}_k\sim dt^2
+M^2\, (\sigma_1^2+\sigma_2^2)
+4t^2\, \sigma_3^2~.
\label{t_2}
\end{eqnarray}
It should be noticed that
at one side of the boundaries the coefficient of $\sigma_1$
vanishes,
while
at the other side it is the coefficient of $\sigma_3$
that vanishes.
The constant $L$ in (\ref{t_1})
is fixed by the ASD condition as 
\begin{eqnarray}
L^2=\frac{3}{\Lambda}\frac{k}{k-2}~.
\label{m^2}
\end{eqnarray}
The asymptotic forms (\ref{t_1}) and (\ref{t_2}) imply that
the metric has an orbifold singularity with angle $2 \pi/(k-2)$
around $\bRP^2$ at $t=t_1$, and extends smoothly over
$\bRP^2$ at $t=t_2$. The principal orbits are
SO(3)/($\bZ_2 \times \bZ_2$) and hence 
the Tod-Hitchin metrics are defined on 
$\bRP^2 \cup
[
(t_1,t_2) \times \mbox{SO(3)}/(\bZ_2 \times \bZ_2)
]
\cup \bRP^2$,
which is topologically equivalent to $S^4$
\cite{Hitchin:a new family}\cite{CGLP}. 
The Taub-NUT de-Sitter metric $g_k$ near the boundary $r=r_+$
coincides with the asymptotic
form  (\ref{t_1}), by setting $t=\int_{r}^{r_+}(1/\sqrt{F(r)})dr$.
However, the metric on the other boundary
$r=n$ is different from (\ref{t_2}), and turns out
to have
the curvature singularity.
The higher order expansions with the initial
conditions (\ref{t_1}) and (\ref{t_2}) 
reveal the further structure of the 
Tod-Hitchin metric.

Using the Einstein equation (see appendix A),
we find the following asymptotic behavior
of the Tod-Hitchin metric in the form (\ref{Bianchi IX})
near the boundary:
\begin{description}
  \item[(1)] Near $t=t_1$
\begin{eqnarray}
a(t) &\sim& \frac{2t}{k-2}+\sum_{j=1}^\infty a_{2j+1} t^{2j+1}~,
\label{near t_1}
\nonumber\\
b(t) &\sim& L + \sum_{j=1}^\infty b_{2j} t^{2j}
+\delta\, t^{k-2}(1+\sum_{n=1}^\infty \delta_{n}t^{n})~,
\label{near t_1}\\
c(t) &\sim& L + \sum_{j=1}^\infty b_{2j} t^{2j}
+\delta\, t^{k-2}(-1+\sum_{n=1}^\infty \hat{\delta}_{n}t^{n})~.\nonumber
\end{eqnarray}
Here the expansion includes one free parameter $\delta$,
and the remaining coefficients are determined by $k$,
$\delta$ and $L$ (see (\ref{m^2})).
In this expansion,
the terms multiplied by $\delta$ represent
the deviation from the biaxial form.
It should be noticed that the deviation is ``small''
because of the presence of the suppression factor 
$t^{k-2}$.\fnote{$\natural$}{In \cite{Hiragane:2003qq}, it was shown that
there exists a similar expansion to (\ref{near t_1}) for
a certain class of higher dimensional Einstein metrics.}
  \item[(2)] Near $t=t_2$
\begin{eqnarray}
a(t)&\sim& M + a_1 t +\sum_{j=2}^\infty a_{j}t^{j}~,\nonumber
\label{near t_2}\\
b(t)&\sim& M -a_1 t + \sum_{j=2}^\infty b_{j}t^{j}~.\\
c(t)&\sim& 2t +\sum_{j=1}^\infty c_{2j+1}t^{2j+1}~.\nonumber
\end{eqnarray}
Here the expansion includes one free parameter $M$,
and the ASD condition requires
\begin{eqnarray}
a_1^2=\frac{1}{4}+\frac{M^2\Lambda}{12}~.
\end{eqnarray}
The remaining coefficients are successively determined.
\end{description}

\begin{figure}
   \parbox{\textwidth}{
   \begin{center}
      \psfig{file=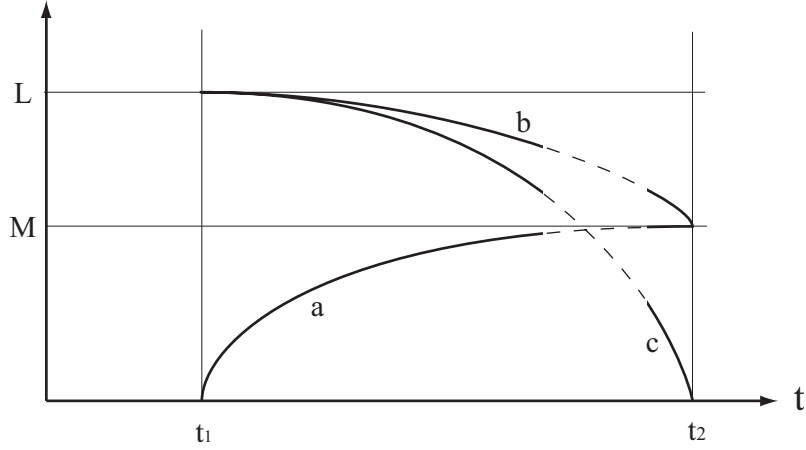,height=60mm}
   \end{center}
   \caption{An illustration of the Tod-Hitchin metric}
\label{fig:fig.2.eps}
}
\end{figure}

The Tod-Hitchin metric corresponds to
that with a certain value $\delta$ in (\ref{near t_1})
or $M$ in (\ref{near t_2});
the determination of these values requires the global
information connecting the local solutions near the boundaries,
which is lacking in our analysis (see Figure \ref{fig:fig.2.eps}).
In particular, for the exact solutions (\ref{exact1})-(\ref{exact2}), the parameters $(\delta,M,\Lambda)$
are given by
\begin{description}
\item [(a)] $k=3$~: \quad $(1,\,\sqrt{3},\,3),~~0 \le t \le \pi/3.$
\item [(b)] $k=4$~: \quad $(3/4,\,1/\sqrt{2},\,6),~~0 \le t \le \pi/4.$ 
\item [(c)] $k=6$~: \quad $(5\sqrt{6}/72,\,1/\sqrt{3},\,3),~~0 \le r \le \infty.$
\item [(d)] $k=8$~: \quad $(63\sqrt{3}/2048,\,\sqrt{3-2\sqrt{2}},\,3),~~\sqrt{2}-1 \le r \le 1.$
\end{description}
When we consider the case with large $k$,
the expansion (\ref{near t_1}) implies
that
the biaxial solutions approximate well
 the Tod-Hitchin
metrics near the boundary $t=t_1$.
We find that
the ASD Taub-NUT de-Sitter solution $g_k$
exactly
reproduces the expansion (\ref{near t_1})
with $\delta=0$.
In the limit $k \rightarrow \infty$, the equation (\ref{near t_1}) yields 
$b(t) \sim c(t)$, which is consistent
with the asymptotic behavior of the Atiyah-Hitchin metric. 
Indeed, the Atiyah-Hitchin metric behaves like the ASD
Taub-NUT metric with exponentially-small corrections
\cite{GM}.

The Atiyah-Hitchin manifold
is identified as
the moduli space
of the three-dimensional $\CN=4$
SU(2) gauge theory\cite{Seiberg}\cite{Seiberg-Witten}.
The vacuum expectation values of bosonic fields of the theory,
three SO(3) scalars $\phi_i$ and one scalar $\sigma$ dual of photon,
parameterize
the Atiyah-Hitchin manifold.
The hyperk\"ahler structure of the Atiyah-Hitchin manifold
ensures the $\CN=4$ supersymmetry.
In the region of large $\left<\phi_i\right>$,
the monopole correction is suppressed and 
the moduli is well approximated by the Taub-NUT geometry
with a negative charge.
On the other hand, near the origin,
the Tod-Hitchin geometry
provides a good approximation even if $k$ is small,
and thus one can expect that 
the gauge theory near the origin of the moduli
is well described by that with the Tod-Hitchin geometry
 as the moduli.
In this approximation, the metric on the moduli becomes simpler
but the gauge theory fails to be supersymmetric.
This is because
the Tod-Hitchin geometry is
not K\"ahler,
while the Atiyah-Hitchin manifold is hyperk\"ahler.
As we have seen,
the Tod-Hitchin geometry converges to the Atiyah-Hitchin manifold
in the  limit, $k\to\infty$ together with $\Lambda\to 0$.
It is interesting 
to consider the gauge theory with the Tod-Hitchin geometry as the moduli and 
to reveal the role of the limit.
In this limit,
the supersymmetry recovers
and
the moduli becomes non compact
by sending the orbifold singularity
 of the Tod-Hitchin geometry to infinity.
On the other hand,
to study the region near the orbifold singularity, 
it will be useful to examine
the theory with the Taub-NUT de Sitter
geometry as the moduli.
These are left for future investigations.

%%%%%%%%%%%%%%%%%%%%%%%%%%%%%%%%%%%%%%%%%%%%%%%%%%%%%%%%%%%%%%%%%%%%%%%
\sect{ Einstein metrics on compact weak $G_2$ manifolds}

In this section we shall describe seven-dimensional geometries
 based on ASD Bianchi IX orbifolds
${\mathcal O_k}$
with the Tod-Hitchin metrics $g^{\rm TH}_k$.
As discussed in the previous section, the Tod-Hitchin
metric is defined on $S^4$ with an orbifold singularity
parameterized by the integer $k$.
However, we shall show that a principal SO(3)
bundle $M_k \rightarrow {\mathcal O_k}$
is actually smooth and the total space $M_k$
admits Einstein metrics of weak holonomy $G_2$.
In this way, 
we obtain an infinite series of seven-dimensional compact
Einstein manifolds.

Let $\phi$ be an SO(3)-connection on $M_k$; it is locally written as
\begin{eqnarray}
\phi = s^{-1} A s + s^{-1} ds~, \quad s \in SO(3).\label{connec}
\end{eqnarray}
Here, $A$ is an so(3)-valued local one-form on 
${\mathcal O_k}$ and $ s^{-1} ds$ is 
regarded as the Maurer-Cartan form. We let $\phi^i$ denote
the component of the connection with respect to
the standard basis $\{E^i \}$ of so(3) which satisfies the Lie bracket relation
$[E^i, E^j] =\epsilon_{i j k} E^k$. The left-invariant
one-forms $\tilde{\sigma_i}$
are defined by $s^{-1}ds= \tilde{\sigma_i} E^i$ and so the equation
(\ref{connec}) may be written as
$\phi^i =s_{ji} A^j+\tilde{\sigma_i}$
by using the adjoint representation
$s^{-1} E^i s = s_{ij} E^j$. Given a metric $\alpha=(\alpha_{ij})$
on SO(3), then the Kaluza-Klein metric on $M_k$ takes the form,
\newcommand\bg{{\boldsymbol g}}
\begin{eqnarray}
\bg_k=\alpha_{ij} \phi^i \phi^j + g^{\rm TH}_k~.\label{KK}
\end{eqnarray}
The Einstein equation
can be solved by imposing the following conditions:
\begin{description}
  \item[(1)] $A^i$ is an SO(3) Yang-Mills instanton
on ${\mathcal O_k}$.
  \item[(2)] The metric $\alpha$ has a diagonal form;
$\alpha=\mbox{diag}(\alpha_1^2,\alpha_2^2,\alpha_3^2)$ where $\alpha_i$ are constants.
\end{description}

The instanton is given by
the self-dual spin connection,
$A^i=-\omega_{0i}- \frac{1}{2}\epsilon_{ijk}\omega_{jk}$~.
Using the explicit formula (\ref{instanton}),
it is written as $A^i= K_i \sigma_i$ with
\begin{eqnarray}
K_1&=& \dot{a}+\frac{-a^2+b^2+c^2}{2 b c}~,\nonumber
\label{ins}\\
K_2&=& \dot{b}+\frac{a^2-b^2+c^2}{2 a c}~,\\
K_3&=& \dot{c}+\frac{a^2+b^2-c^2}{2 a b}~.\nonumber
\end{eqnarray}
Thus, the seven-dimensional Einstein equations 
with cosmological constant $\lambda$ are equivalent to
\begin{eqnarray}
\frac{\alpha_1^4-(\alpha_2^2-\alpha_3^2)^2}{2 \alpha_1^2
\alpha_2^2 \alpha_3^2}+\left(\frac{\Lambda}{3} \right)^2 \alpha_1^2=\lambda,
\quad \Lambda-\frac{1}{2}\left(\frac{\Lambda}{3} \right)^2
(\alpha_1^2+\alpha_2^2+\alpha_3^2)=\lambda,
\end{eqnarray}
and the two equations
with cyclic permutation
of
$\alpha_1, \alpha_2, \alpha_3$. 
These can be solved easily, and one has
two solutions,
\begin{eqnarray}
\alpha=\beta_{\ell}\,\mbox{diag}(1,1,1), \quad
\beta_{\ell}=\frac{3}{\ell \Lambda} \label{sol}
\end{eqnarray}
with $\lambda=\Lambda \frac{2\ell-1}{2 \ell}$~($\ell$=1 or $5$). 
Using the right-invariant one-forms $\hat{\sigma_i}$
($s ds^{-1}=\hat{\sigma_i} E^i$) and the Tod-Hitchin metric in the form
(\ref{Bianchi IX}), we find
 two types of seven-dimensional Einstein metrics;
\begin{eqnarray}
\bg_k^{(\ell)}=dt^2+a^2(t)\sigma_1^2+b^2(t)\sigma_2^2+c^2(t)\sigma_3^2
+\beta_{\ell}(K_{i}(t) \sigma_i-\hat{\sigma_i})^2~.\label{sol2}
\label{sasaki}
\end{eqnarray}

The conditions (1) and (2) also induce a $G_2$-structure
on $M_k$ as follows:
Recall that the $G_2$-structure is characterized by a
global one-form $\omega$, which is written locally as
\begin{eqnarray}
\omega&=&\theta^1 \wedge \theta^2 \wedge \theta^3+\theta^1 \wedge(
\theta^4 \wedge \theta^5+\theta^6 \wedge \theta^7)~\nonumber
\\
&+&\theta^2 \wedge(
\theta^4 \wedge \theta^6+\theta^7 \wedge \theta^5)
+\theta^3 \wedge(
\theta^4 \wedge \theta^7+\theta^5 \wedge \theta^6)
~,\label{G2}
\end{eqnarray}
where \{$\theta^{\alpha}; \alpha=1,2, \dots 7$ \} is a 
fixed orthonormal basis of the seven-dimensional metric $\bg_{diag}$
 (see appendix C).
The condition of weak holonomy $G_2$ is defined by
$d\omega=c *\omega$ where $*$ is the Hodge star operation associated
to $\bg_{diag}$ and
$c$ is a constant. Under (1) and (2), the weak $G_2$ condition
reproduces the metric (\ref{sol2}). The holonomy group
$\mbox{Hol}(\bar \bg_k^{(\ell)})$ of the metric cone 
$( C(M_k),\bar \bg_k^{(\ell)})=(R_{+} \times M_k, d\tau^2+\tau^2 \bg_k^{(\ell)})$
is contained in Spin(7) \cite{Friedrich:1995dp}\cite{BG}: 
\begin{description}
  \item[(A)] $\mbox{Hol}(\bar \bg_k^{(1)})=$ Sp(2) $\subset $ Spin(7) and 
$(M_k, \bg_k^{(1)})$ is a
3-Sasakian manifold.
%\fnote{$\sharp$}{
%Grove, Wilking and Ziller proved that for
%odd $k$ the 3-Sasakian manifold $M_k$
%is a rational homology seven-sphere
%with non-zero torsion depending on $k$.
%This statement is
% written 
%in  example 22 \cite{Boyer-Galicki}
%as a private communication with Ziller.
%We could not find their paper unfortunately.
%}
  \item[(B)] $\mbox{Hol}(\bar \bg_k^{(5)}))$ = Spin(7) and
$(M_k, \bg_k^{(5)}))$ is a
proper $G_2$ manifold.
\end{description}

We now proceed to a discussion of the metric singularities.
The orbifold singularity of the base space ${\mathcal O_k}$
emerges at the boundaries
where a certain component of the metric vanishes.
To understand the effect of this singularity in the total space $M_k$,
it is useful to see the behavior of the metric $\bg_k^{(\ell)}$ with
weak holonomy $G_2$
near boundaries. From (\ref{near t_1}) and (\ref{near t_2}),
putting $\Omega(k)=k^2+(k-2)^2$ we find
\begin{eqnarray}
\bg_k^{(\ell)} &\rightarrow& 
dt^2+\frac{4 t^2}{\Omega^2(k)} ((k-2)\sigma_1+k \hat{\sigma_1})^2 \nonumber\\
&&+
\frac{\ell\beta_{\ell}k}{k-2}(\sigma_2^2+\sigma_3^2)+\beta_{\ell}
(\hat{\sigma_2}^2+\hat{\sigma_3}^2)+\frac{\beta_{\ell}}{(k-2)^2}
(k \sigma_1-(k-2)\hat{\sigma_1})^2
\end{eqnarray}
for $t \rightarrow t_1$, and
\begin{eqnarray}
\bg_k^{(\ell)} &\rightarrow& 
dt^2+\frac{t^2}{25}(\sigma_3+3 \hat{\sigma_3})^2 \nonumber\\
&&+
M^2(\sigma_1^2+\sigma_2^2)+\beta_{\ell}
(\hat{\sigma_1}^2+\hat{\sigma_2}^2)+\beta_{\ell}
(3\sigma_3-\hat{\sigma_3})^2
\end{eqnarray}
for $t \rightarrow t_2$. These expressions correspond to
the asymptotic forms (\ref{t_1}) and (\ref{t_2}) of the Tod-Hitchin
metric. An important difference is that the collapsing circle
is twisted by the fiber SO(3), which allows us to resolve the orbifold
singularity of ${\mathcal O_k}$ as shown below.
Let us represent the invariant
one-forms $\sigma_i, \hat{\sigma_j}$ in terms of Euler's angles:
\begin{eqnarray}
\sigma_1 &=& d\psi+\cos\theta d\phi, \quad \hat{\sigma}_1=-d\hat{\phi}
-\cos\hat{\theta}d\hat{\psi}, \nonumber\\ 
\sigma_2 &=& \cos \psi d\theta+\sin\psi \sin\theta d\phi, \quad
\hat{\sigma}_2=-\cos\hat{\phi}d\hat{\theta}
-\sin\hat{\phi}\sin\hat{\theta}d\hat{\psi}, \nonumber\\
\sigma_3 &=& -\sin\psi d\theta+\cos\psi \sin\theta d\phi, \quad
\hat{\sigma}_3=-\sin\hat{\phi}d\hat{\theta}+
\cos\hat{\phi}\sin\hat{\theta}d\hat{\psi}~.
\end{eqnarray}
The following transformation
\begin{eqnarray}
\eta=\frac{2}{\Omega(k)}((k-2)\psi-k\hat{\phi}), \quad
\chi=k \psi+(k-2)\hat{\phi}~,
\label{angle}
\end{eqnarray}
yields
\begin{eqnarray}
\bg_k^{(\ell)} &\rightarrow& 
dt^2+t^2\left(d\eta+\frac{2(k-2)}{\Omega(k)}\cos \theta d\phi
-\frac{2 k}{\Omega(k)}\cos\hat{\theta}d\hat{\psi} \right)^2
\nonumber\\
&&+
\frac{\ell\beta_{\ell}k}{k-2}(d\theta^2+\sin^2\theta d\phi^2)+\beta_{\ell}
(d\hat{\theta}^2+\sin^2\hat{\theta}d\hat{\psi}^2)
\nonumber\\
&&+\frac{\beta_{\ell}}{(k-2)^2}
(d\chi+k \cos \theta d\phi +(k-2)\cos\hat{\theta}d\hat{\psi})^2
\end{eqnarray}
for $t \rightarrow t_1$. From (\ref{angle}) we have $d\eta \wedge d\chi
=2(d\psi \wedge \hat{\phi})$. It follows that
one can adjust the ranges of
the new angles as
$0 \le \eta < 2 \pi$, $0 \le \chi < 4 \pi$
since Euler's angles have
 the ranges $0 \le \psi < 2 \pi$, 
$0 \le \hat{\phi} < 2 \pi$. Thus, the metric $\bg_k^{(\ell)}$ extends 
smoothly over the circle bundle $T^{k,k-2}$ with the squashed metric
\begin{eqnarray}
g_{bolt}=\frac{\ell k}{k-2}(d\theta^2+\sin^2\theta d\phi^2)+
d\hat{\theta}^2+\sin^2\hat{\theta}d\hat{\psi}^2\nonumber\\
+\frac{1}{(k-2)^2}
(d\chi+k \cos \theta d\phi +(k-2)\cos\hat{\theta}d\hat{\psi})^2
\end{eqnarray}
at the boundary $t=t_1$. Also, similar arguments show that the metric
extends over $T^{3,1}$ at $t=t_2$.

%%%%%%%%%%%%%%%%%%%%%%%%%%%%%%%%%%%%%%%%%%%%%%%%%%%%%%%%%%%%%%%%%%%%%%%
\sect{Geodesics on weak $G_2$ manifolds}
In this section,
 we consider a Hamiltonian formulation
describing geodesics on the weak $G_2$ manifold $M_k$.
The phase space is the cotangent bundle $T^{*}M_k$ with
coordinates $(x^\alpha)=(t, \theta, \phi, \psi, \hat{\theta},
\hat{\phi},\hat{\psi})$ and their conjugate momenta $(p_\alpha)$.
The equations for geodesic flow are the canonical equations on
$T^{*}M_k$ with Hamiltonian $H=\frac{1}{2}g^{\alpha \beta}p_{\alpha}
p_{\beta}$. Using the metric (\ref{sasaki}), we may write explicitly as
\begin{eqnarray}
H&=&
\frac{1}{2}p_t^2+\frac{1}{2}\left(\frac{L_1^2}{a^2} +
 \frac{L_2^2}{b^2} + \frac{L_3^2}{c^2}\right)
+\frac{1}{2 \beta_{\ell}}(\hat{R}_1^2 +\hat{R}_2^2+\hat{R}_3^2)
\nonumber\\
&+& 
\frac{1}{2} \left(
\frac{K_1^2 \hat{R}_1^2}{a^2}+\frac{K_2^2 \hat{R}_2^2}{b^2} 
  + \frac{K_3^2 \hat{R}_3^2}{c^2}\right)  
+\frac{K_1 L_1 \hat{R}_1}{a^2}+\frac{K_2 L_2 \hat{R}_2}{b^2}+
\frac{K_3 L_3 \hat{R}_3}{c^2}.
\end{eqnarray}
The functions $L_i$ and $\hat{R}_j$ are canonically conjugate to
$\sigma_i$ and $\hat{\sigma_j}$, respectively:
\begin{eqnarray}
L_1 &=& p_{\psi}~, \nonumber \\
L_2 &=& -\cot \theta \sin \psi p_{\psi}+\cos \psi p_{\theta}+
\frac{\sin \psi}{\sin\theta} p_{\phi}~, \nonumber \\
L_3 &=& -\cot \theta \cos \psi p_{\psi}-\sin \psi p_{\theta}+
\frac{\cos \psi}{\sin\theta} p_{\phi}~, \nonumber \\
\hat{R}_1 &=& -p_{\hat{\phi}}~, \nonumber \\
\hat{R}_2 &=& \cot \hat{\theta} \sin \hat{\phi} p_{\hat{\phi}}
-\cos \hat{\phi} p_{\hat{\theta}}-
\frac{\sin \hat{\phi}}{\sin\theta} p_{\hat{\psi}}~, \nonumber \\
\hat{R}_3 &=& -\cot \hat{\theta} \cos \hat{\phi} p_{\hat{\phi}}
-\sin \hat{\phi} p_{\hat{\theta}}+
\frac{\cos \hat{\phi}}{\sin\hat{\theta}} p_{\hat{\psi}}~, 
\end{eqnarray}
which satisfy the SO(3) $\times$ SO(3) relations,
$\{ L_i , L_j \}=-\epsilon_{ijk} L_k$ and $\{\hat{R}_i, \hat{R}_j \}
=-\epsilon_{ijk} \hat{R}_k$. We also introduce functions
$\hat{L}_i$ and $R_j$ by exchanging Euler's angles, 
$(\theta, \phi, \psi) \leftrightarrow 
(\hat{\theta},\hat{\phi},\hat{\psi})$. Then, one can easily show that
they express the isometry SO(3) $\times$ SO(3) of the metric;
$\{L_i, R_j \}=\{\hat{L}_i, \hat{R}_j \}=0$ and hence
$\{H, \hat{L}_i \}=\{H, R_j \}=0$. 
It should be noticed that in general neither $L_i$ nor $\hat{R}_j$
are conserved,
although $\sum_i L_i^2 = \sum_i R_i^2$ and 
$\sum_i \hat{L}_i^2 = \sum_i \hat{R}_i^2$
are conserved quantities, the second Casimir.
 The relation between
$L_i~ (\hat{L_i})$ and $R_i~ (\hat{R}_i)$ corresponds to
the relation between left and right actions of SO(3).
The Hamiltonian equations $\frac{df}{d\tau}=\{ f, H \}$ are
\begin{eqnarray}
\frac{dL_1}{d\tau} &=& \left( \frac{1}{c^2}-\frac{1}{b^2} \right)
L_2 L_3- \frac{K_2}{b^2} L_3 \hat{R}_2+\frac{K_3}{c^2} L_2 \hat{R}_3~,
\nonumber\\
\frac{dL_2}{d\tau} &=& \left( \frac{1}{a^2}-\frac{1}{c^2} \right)
L_3 L_1- \frac{K_3}{c^2} L_1 \hat{R}_3+\frac{K_1}{a^2} L_3 \hat{R}_1~,
\nonumber\\
\frac{dL_3}{d\tau} &=& \left( \frac{1}{b^2}-\frac{1}{a^2} \right)
L_1 L_2- \frac{K_1}{a^2} L_2 \hat{R}_1+\frac{K_2}{b^2} L_1 \hat{R}_2~,
\label{L} 
\end{eqnarray}
and
\begin{eqnarray}
\frac{d\hat{R}_1}{d\tau} &=&
\left(\left( \frac{K_3}{c} \right)^2- \left(\frac{K_2}{b} \right)^2
\right)
\hat{R}_2 \hat{R}_3- \frac{K_2}{b^2} \hat{R}_3 L_2+
\frac{K_3}{c^2} \hat{R}_2 L_3~,
\nonumber\\
\frac{d\hat{R}_2}{d\tau} &=& 
\left( \left(\frac{K_1}{a} \right)^2-\left(\frac{K_3}{c} \right)^2
\right)
\hat{R}_3 \hat{R}_1- \frac{K_3}{c^2} \hat{R}_1 L_3+
\frac{K_1}{a^2} \hat{R}_3 L_1~,
\nonumber\\
\frac{d\hat{R}_3}{d\tau} &=& 
\left( \left(\frac{K_2}{b} \right)^2-\left(\frac{K_1}{a} \right)^2
 \right)
\hat{R}_1 \hat{R}_2- \frac{K_1}{a^2} \hat{R}_2 L_1+
\frac{K_2}{b^2} \hat{R}_1 L_2~ \label{R}
\end{eqnarray}
together with
\begin{eqnarray}
\frac{dt}{d\tau} &=& p_t~, \nonumber\\
\frac{d p_t}{d\tau} &=& 
\frac{\dot{a}}{a^3} L_1^2+\frac{\dot{b}}{b^3} L_2^2+
\frac{\dot{c}}{c^3}L_3^2 \nonumber \\
&&- \frac{K_1}{a} \left( \frac{\dot{K_1}}{a}-\frac{K_1 \dot{a}}{a^2}
\right) \hat{R}_1^2
- \frac{K_2}{b} \left( \frac{\dot{K_2}}{b}-\frac{K_2 \dot{b}}{b^2}
\right) \hat{R}_2^2
- \frac{K_3}{c} \left( \frac{\dot{K_3}}{c}-\frac{K_3 \dot{c}}{c^2}
\right) \hat{R}_3^2 \nonumber \\
&&- \left( \frac{\dot{K_1}}{a^2}-2\frac{K_1 \dot{a}}{a^3}
\right) L_1 \hat{R}_1
- \left( \frac{\dot{K_2}}{b^2}-2\frac{K_2 \dot{b}}{b^3}
\right) L_2 \hat{R}_2
- \left( \frac{\dot{K_3}}{c^2}-2\frac{K_3 \dot{c}}{c^3}
\right) L_3 \hat{R}_3~. \label{time}
\end{eqnarray}

This system may be regarded as an interacting rigid body system
with angular momenta $L_i$ and $\hat{R}_j$. The moments 
of inertia are given by $(I_i)=(a, b, c)$
and
$(\hat{I}_i)=(a/K_1, b/K_2, c/K_3)$,
which have a non-trivial time dependence through the equation
(\ref{time}). When we put $K_i=0$, then the interaction between
$L_i$ and $\hat{R}_j$ vanishes. Thus, the angular momenta $\hat{R}_j$
are constants, and the remaining equations (\ref{L}) and (\ref{time})
 describe the geodesics on the Tod-Hitchin manifold
\cite{Hitchin:a new family}\cite{GM}\cite{BM}.

As a special solution, consider the case $L_2=\hat{R}_2=0$ 
in the equations (\ref{L})-(\ref{time}). Then, the angular momenta
$(L_1, L_3)$ and $(\hat{R}_1, \hat{R}_3)$ are constants. If we can
find a parameter $t_0$ such that $a(t_0)=c(t_0)$, we have
$\frac{d L_2}{d\tau}=\frac{d \hat{R}_2}{d\tau}=0$ after setting
\begin{eqnarray}
&K_3&(t_0)L_1 \hat{R}_3-K_1(t_0)L_3 \hat{R}_1=0~, \nonumber\\
(&K_1^2&(t_0)-K_3^2(t_0))\hat{R}_3 \hat{R}_1-K_3(t_0)L_3 \hat{R}_1
+K_1(t_0)L_1 \hat{R}_3=0~.\label{const1}
\end{eqnarray}
In fact, one can show that the parameter $t_0$ exists from the
behavior of the Painlev\'e VI solution (see Figure 2).
Finally, the equation $p_t =0$ requires the further constraint
for the angular momenta:
\begin{eqnarray}
\frac{\dot{a}}{a} &L_1^2&+\frac{\dot{c}}{c}L_3^2+
K_1\left( \frac{a \Lambda}{3}+K_1 \frac{\dot{a}}{a} \right)\hat{R}_1^2
+
K_3\left( \frac{a \Lambda}{3}+K_3 \frac{\dot{c}}{c} \right)\hat{R}_3^2
\nonumber \\
&+&\left( \frac{a \Lambda}{3}+2 K_1 \frac{\dot{a}}{a} \right)
L_1 \hat{R}_1 +
\left( \frac{a \Lambda}{3}+2 K_3 \frac{\dot{c}}{c} \right)
L_3 \hat{R}_3 =0~,\label{const2}
\end{eqnarray}
where we have used an identity $\dot{K}_1=\dot{K}_3=-a \Lambda/3$
at $a=c$. If we consider the case $\hat{R}_1=\hat{R}_3=0$,
the equation (\ref{const1}) is automatically satisfied, and
(\ref{const2}) yields $(L_1/L_3)^2=-(\dot{c}/\dot{a})(t_0)$\cite{BM}.
As a result, we find a class of geodesics on $M_k$.
For cases $k=3,4,6$ and 8 given by
(\ref{exact1})-(\ref{exact2}), the solutions
are summarized as follows: 
\begin{description}
\item [(a)] $k=3$~:~~$t_0=\pi/6$\\
$\frac{L_1}{L_3}= \pm 1,~~\hat{R}_1=\hat{R}_3=0,$\\ 
$\frac{L_1}{\hat{R}_3}=\frac{\hat{R}_1}{\hat{R}_3}-\sqrt{3},
~~\frac{L_3}{\hat{R}_3}=1+\sqrt{3},$\\
$\frac{L_1}{\hat{R}_1}=-2/(1+\sqrt{3})$~ and
$-13/(3+4\sqrt{3}),~~L_3=\hat{R}_3=0.$
\item [(b)] $k=4$~:~~$t_0=\pi/6$\\
$\frac{L_1}{L_3}= \pm 2,~~\hat{R}_1=\hat{R}_3=0,$\\ 
$\frac{L_1}{\hat{R}_3}=-\sqrt{3}~\frac{\hat{R}_1}{\hat{R}_3},
~~\frac{L_3}{\hat{R}_3}=\sqrt{3}/2,$\\
$\frac{L_1}{\hat{R}_1}=\sqrt{3}$~ and
$-4\sqrt{3}/3,~~L_3=\hat{R}_3=0.$
\item [(c)] $k=6$~:~~$r_0=2^{1/3}+2^{-1/3} \cong 2.05$\\
$\frac{L_1}{L_3} \cong \pm 1.92,~~\hat{R}_1=\hat{R}_3=0,$\\
$\frac{L_1}{\hat{R}_1} \cong -1.71$ and $-1.28$,
~~$L_3=\hat{R}_3=0,$\\
$\frac{L_3}{\hat{R}_3} \cong 0.95$ and 1.06,
~~$L_1=\hat{R}_1=0.$
\item [(d)] $k=8$~:~~$r_{0} \cong 0.55$\\ 
$\frac{L_1}{L_3} \cong \pm 2.21,~~\hat{R}_1=\hat{R}_3=0,$\\
$\frac{L_1}{\hat{R}_1} \cong -1.15,~~
\frac{L_3}{\hat{R}_1} \cong \pm 0.50,~~
\frac{\hat{R}_3}{\hat{R}_1} \cong \pm 0.52,$\\
$\frac{L_1}{\hat{R}_1} \cong -1.46$ and $-1.15$,
~~$L_3=\hat{R}_3=0,$\\
$\frac{L_3}{\hat{R}_3} \cong 0.97$ and 1.03,
~~$L_1=\hat{R}_1=0.$
\end{description}

%%%%%%%%%%%%%%%%%%%%%%%%%%%%%%%%%%%%%%%%%%%%%%%%%%%%%%%%%%%%%%%%%%%%
\sect{M-theory on AdS$_4\times M_k$}

We have constructed infinitely many compact
Einstein
manifolds $M_k$,
 which are 3-Sasakian manifolds for $\ell=1$
and proper weak $G_2$ manifolds for $\ell=5$.
The orbifold singularity of the
Tod-Hitchin geometry
has been resolved by having additional dimensions,
so that we can expect the resolution of the orbifold singularity
in the moduli
by adding scalars in the corresponding gauge theory.
The resulting seven-dimensional manifolds $M_k$
admit 3-Sasakian or proper weak $G_2$
structures, and thus the gauge theories are $\CN=3$ supersymmetric
for $\ell=1$, while $\CN=1$ supersymmetric for $\ell=5$.
It was shown that the manifold $M_3(\ell=1)=N^{0,1,0}$
appears as the moduli space of an $\CN=3$
gauge theory \cite{N2}.
We expect that the seven-dimensional manifolds $M_k$
with general $k$ also
emerge as the moduli spaces
of three-dimensional $\CN=3$ or $\CN=1$
supersymmetric gauge theories.
It is interesting to achieve this and 
to reveal the role of $k$
from the viewpoint of gauge theories.
Leaving this interesting issue as a future problem,
in this section
we consider M-theory on AdS$_4\times M_k$,
and apply the AdS/CFT correspondence.

Using the
3-Sasakian or proper weak $G_2$
manifolds $M_k$,
one can construct supersymmetric M-theory solutions,
AdS$_4\times M_k$,
which are AdS/CFT dual to three-dimensional superconformal field theories.
The isometry of $M_k$ corresponds to the  global symmetry of 
the dual superconformal field theories,
including the R-symmetry.
The manifolds $M_k$ contain S$^7$, $N^{0,1,0}$ and 
squashed S$^7$ ($\tilde S^7$) as
special homogeneous cases; $M_3(\ell=1)$, $M_4(\ell=1)$ and $M_3(\ell=5)$,
respectively.
For these cases,
the dual three-dimensional gauge theories
which  flow to the superconformal field theories
 at the IR
are
the $\CN=8$ gauge theory without flavor \cite{Maldacena} 
for S$^7$ with SO(8) isometry,
the $\CN=3$ gauge theory with SU(3) flavor
\cite{N2}\cite{N^{0,1,0}} 
for $N^{0,1,0}$ with SU(3)$\times$SU(2) isometry.
The squashed S$^7$ admits SO(5)$\times$SO(3) isometry
so that the dual theory is expected to be
$\CN=1$ gauge theory with SO(5)$\times$SO(3) flavor.
For generic $k$,
because the metrics on $M_k$ admit
SO(3)$\times$SO(3) isometry as shown in section 4,
the gauge theories which flow to the superconformal field theories
 at the IR
are an $\CN=3$ gauge theory with  SO(3) flavors for $\ell=1$,
and an $\CN=1$ gauge theory with  SO(3)$\times$SO(3) flavors for $\ell=5$.
Since it is not easy to extract the Kaluza-Klein spectrum
on $M_k$ as is expected from the analysis in section 4,
we assume this correspondence here.
The UV limit of the theory
is  described by $\bR^{1,2}\times C(M_k)$,
where $C(M_k)$ stands for the cone over $M_k$.
The cone metric 
are 
hyperk\"ahler for $\ell=1$ and Spin(7) for $\ell=5$.
For the
homogeneous cases 
S$^7$, 
$N^{0,1,0}$ and 
 $\tilde S^7$,
the holographic RG-flows which interpolate
$\bR^{1,2}\times C(M_k)$ at UV and AdS$_4\times M_k$ at IR
are examined in \cite{GNS}.
For general $k$,
the brane solution
which describes the holographic RG-flow
from $\bR^{1,2}\times C(M_k)$ at UV to AdS$_4\times M_k$ at IR
is
\begin{eqnarray}
g_{11}=H^{-\frac{2}{3}}g_{\bR^{1,2}}+H^{\frac{1}{3}}\bar \bg_{k}^{(\ell)}~,
~~~~F={\rm dvol}(\bR^{1,2})\wedge dH^{-1} ~,~~~~
H=1+\left(\frac{a}{r}\right)^6,
\label{brane solution}
\end{eqnarray}
where $a=(2^5\pi^2N)^{\frac{1}{6}}\ell_P$ and 
$\bar\bg_{k}^{(\ell)}=dr^2+r^2 \bg_k^{(\ell)}$.
This corresponds to $N$ coincident M2-branes at $r=0$.
For small $r$, the brane solution (\ref{brane solution})
reduces to the product metric of $M_k$ with cosmological constant $1/a^2$
and AdS$_4$ with $4/a^2$, and
the four-form strength 
$F=6$\,dvol(AdS$_4$)$/a$.
On the other hand, for large  $r$,
(\ref{brane solution}) approaches the product metric of
$\bR^{1,2}$ and $C(M_k)$ without the four-form strength.
It is interesting to examine
the limit, $k\to\infty$ together with $\Lambda\to 0$,
in which the four-dimensional base space, Tod-Hitchin geometry,
converges to the Atiyah-Hitchin hyperk\"ahler manifold $M_{AH}$.
The limit $\Lambda\to 0$ corresponds to the limit
 $a\to \infty$, because
the cosmological constant $\lambda=\Lambda\frac{2\ell-1}{2\ell}$
of $M_k$
is now ${1}/{a^2}$.
In this limit, (\ref{brane solution})
approaches
the metric on
$\bR^{1,3}\times \bR^{3}/\bZ_2\times M_{AH}$
without the four-form strength
because $M_k$ reduces to $\bR^{3}/\bZ_2\times M_{AH}$.
Apart from the $\bZ_2$ factor,
this solution can be regarded as an orientifold 6-plane of 
the IIA superstring theory,
and thus the $g_{11}$ provides an approximation of
the orientifold plane.

Infinitely many inhomogeneous Einstein metrics on compact manifolds
are derived from Kerr de-Sitter black holes
as the Page limit in 
\cite{HSY}\cite{Gibbons:2004uw}\cite{Gibbons:2004js},
and those with a Sasaki structure found in
\cite{Gauntlett:2004yd}
 as the Sasaki-Einstein twist in \cite{HSY2}.
It is interesting to consider the black hole solutions
corresponding to $M_k$ constructed in this paper.
We have discussed the holographic RG-flow
from $\bR^{1,2}\times C(M_k)$ to AdS$_4\times M_k$.
In \cite{AR}, a transition from AdS$_4\times \tilde S^7$ to AdS$_4\times S^7$
is discussed.
It is expected that there is a similar transition
from  AdS$_4\times M_k(\ell=5)$ to AdS$_4\times M_k(\ell=1)$.
We leave these issues for future investigations.

\newpage

\noindent
{\bf Note added}:
After submitting this paper to e-print archives,
we received from K. Galicki
the draft \cite{Ziller} of a talk given by W. Ziller,
which is refereed in
\cite{Boyer-Galicki}.
In the draft,
Grove, Wilking and Ziller proved that
3-Sasakian orbifolds $M_k(\ell=1)$
corresponding to 
AdS Bianchi IX orbifolds ${\mathcal O}_k$ with the Tod-Hitchin metrics
are manifolds with the following properties:
(a) for odd $k$, they have the same cohomology ring as an S$^3$-bundle
over S$^4$, 
(b) for even $k$, they have the same cohomology ring as a general
Aloff Wallach space, 
(c) in both cases, it carries an invariant cohomogeneity one structure by
S$^3\times$S$^3$.
In addition, we were informed
by K. Galicki that the proper weak $G_2$ orbifolds $M_k(\ell=5)$
can be also made smooth by the method of K. Galicki and S. Salamon
\cite{BG}.

Our study 
provides 
a concrete procedure to resolve orbifold singularities
which is familiar to physicists,
and the explicit forms of the 3-Sasakian and proper weak $G_2$ metrics.

\section*{Acknowledgements}
The authors thank Yoshitake Hashimoto
for useful discussions,
and Krzysztof Galicki for correspondence.
Y.Y. is grateful to Gary Gibbons for his kind hospitality
and useful discussions
during his stay at DAMTP in University of Cambridge.
This paper is supported by the 21 COE program
``Constitution of wide-angle mathematical basis focused on knots".
Research of Y.Y. is supported  in part by the Grant-in
Aid for scientific Research (No.~14540073 and No.~14540275)
from Japan Ministry of Education.

\vspace{4mm}

%%%%%%%%%%%%%%%%%%%%%%%%%%%%%%%%%%%%%%%%%%%%%%%%%%%%%%%%%%%%%%%%%%%%%%%%
\appendix
\sect{four-dimensional ASD Einstein manifolds}
The Bianchi IX metric is of the form
\begin{eqnarray}
g=dt^2
+a^2(t)\sigma_1^2
+b^2(t)\sigma_2^2
+c^2(t)\sigma_3^2~,
\end{eqnarray}
where $\sigma_i$ are left-invariant one-forms on SO(3)~,
\begin{eqnarray}
d\sigma_i=-\frac{1}{2}\epsilon_{ijk}\sigma_j\wedge\sigma_k.
\end{eqnarray}
Defining vielbein
\begin{eqnarray}
e^0=dt,~~~
e^1=a\sigma_1,~~~
e^2=b\sigma_2,~~~
e^3=c\sigma_3, \label{orth}
\end{eqnarray}
one evaluates the spin connection as
\begin{eqnarray}
\omega_{01}&=&-\frac{\dot a}{a}~e^1~,~~~~
\omega_{12}=-\frac{a^2+b^2-c^2}{2abc}~e^3~,
\nonumber\\
\omega_{02}&=&-\frac{\dot b}{b}~e^2~,~~~~
\omega_{31}=-\frac{a^2-b^2+c^2}{2abc}~e^2~,
\nonumber\\
\omega_{03}&=&-\frac{\dot c}{c}~e^3~,~~~~
\omega_{23}=-\frac{-a^2+b^2+c^2}{2abc}~e^1~.
\label{instanton}
\end{eqnarray}
The Einstein equations 
$R_{\alpha \beta}=\Lambda \delta_{\alpha \beta}$
are given by
\begin{eqnarray}
\frac{\ddot a}{a}
&+&\frac{\ddot b}{b}
+\frac{\ddot c}{c}+\Lambda=0~,\nonumber\\
\frac{\ddot a}{a}
&+&\frac{\dot a}{a}
\left(
 \frac{\dot b}{b} + \frac{\dot c}{c}
 \right)
-\frac{a^4-(b^2-c^2)^2}{2a^2b^2c^2}+\Lambda=0
~,\nonumber\\
\frac{\ddot b}{b}
&+&\frac{\dot b}{b}
 \left(
 \frac{\dot a}{a} + \frac{\dot c}{c}
 \right)
-\frac{b^4-(a^2-c^2)^2}{2a^2b^2c^2}+\Lambda=0
~,\nonumber\\
\frac{\ddot c}{c}
&+&\frac{\dot c}{c}
 \left(
 \frac{\dot a}{a} + \frac{\dot b}{b}
 \right)
-\frac{c^4-(a^2-b^2)^2}{2a^2b^2c^2}+\Lambda=0
~.
\end{eqnarray}
The ASD condition further requires the following equations:
\begin{eqnarray}
\frac{\ddot a}{a}~
&+&\left(
 C \frac{\dot b}{b}+B \frac{\dot c}{c}-\frac{\dot a}{bc}
 \right)+\frac{\Lambda}{3}=0~,\nonumber\\
\frac{\ddot b}{b}~
&+&\left(
 C \frac{\dot a}{a}+A \frac{\dot c}{c}-\frac{\dot b}{ac}
 \right)+\frac{\Lambda}{3}=0~,\nonumber\\
\frac{\ddot c}{c}
&+&\left(
 B \frac{\dot a}{a}+A \frac{\dot b}{b}-\frac{\dot c}{ab}
 \right)+\frac{\Lambda}{3}=0~,\nonumber\\ 
\frac{\dot a \dot b}{a b}~
&-&
\frac{a^4+b^4-3c^4+2(-a^2b^2+b^2c^2+a^2c^2)}{4a^2b^2c^2}
+\left(
 B\frac{\dot a}{a}+A\frac{\dot b}{b}-\frac{\dot c}{ab}
\right)
+\frac{\Lambda}{3}=0~,
\nonumber\\
\frac{\dot a \dot c}{a c}~
 &-&
\frac{a^4-3b^4+c^4+2(a^2b^2+b^2c^2-a^2c^2)}{4a^2b^2c^2}
+\left(
 C\frac{\dot a}{a}+A\frac{\dot c}{c}-\frac{\dot b}{ac}
\right)
+\frac{\Lambda}{3}=0~,\nonumber\\
\frac{\dot b \dot c}{b c}~
&-& 
\frac{-3a^4+b^4+c^4+2(a^2b^2-b^2c^2+a^2c^2)}{4a^2b^2c^2}
+\left(
 C\frac{\dot b}{b}+B\frac{\dot c}{c}-\frac{\dot a}{bc}
\right)
+\frac{\Lambda}{3}=0~,
\end{eqnarray}
where
\begin{eqnarray}
A=\frac{-a^2+b^2+c^2}{2abc},~~~
B=\frac{a^2-b^2+c^2}{2abc},~~~
C=\frac{a^2+b^2-c^2}{2abc}.
\end{eqnarray}

%%%%%%%%%%%%%%%%%%%%%%%%%%%%%%%%%%%%%%%%%%%%%%%%%%%%%%%%%%%%%
\sect{Tod-Hitchin metric}

Tod \cite{Tod} and Hitchin \cite{Hitchin:twistor}
\cite{Hitchin:a new family}
studied the Bianchi IX metric written in the form
\begin{eqnarray}
g^{\rm TH} = 
H(x)  \left(
 \frac{dx^2}{x  (1-x)} +
\frac{\sigma_1^2}{\Omega_1(x)^2} +
 \frac{(1-x)  \sigma_2^2}{\Omega_2(x)^2}
 + \frac{x  \sigma_3^2}{\Omega_3(x)^2}
\right) ~.
\label{TH}
\end{eqnarray}
They showed that 
$g^{\rm TH}$ gives an ASD Einstein metric with positive cosmological constant
 if the functions $\Omega_i$ satisfy
a set of first order equations
\begin{eqnarray}
\Omega_1'= -\frac{\Omega_2  \Omega_3}{x  (1-x)} ~,~~~~
\Omega_2' = -\frac{\Omega_3  \Omega_1}{x} ~,~~~~
\Omega_3' = -\frac{\Omega_1  \Omega_2}{1-x} ~,
\label{Omega}
\end{eqnarray}
where a prime denotes a derivative with respect to $x$,
and the conformal factor $H$ is given by 
\begin{eqnarray}
H=
-\frac{8x  \Omega_1^2  \Omega_2^2  \Omega_3^2 
+ 2\Omega_1 \Omega_2  \Omega_3  
\left\{
 x  (\Omega_1^2+\Omega_2^2) -(1-4\Omega_3^2)(\Omega_2^2 -(1-x) \Omega_1^2)
\right\}
}{
4
\left\{
 x  \Omega_1  \Omega_2 + 2\Omega_3 \left(\Omega_2^2 - (1-x)  \Omega_1^2\right)
\right\}^2} ~.
\label{H}
\end{eqnarray}
Writing the functions $\Omega_i^2$
in terms of  $y(x)$
as
\begin{eqnarray}
\Omega_1^2 &=& \frac{(y-x)^2  y  (y-1)}{x  (1-x)}  \left(
z-\frac{1}{2(y-1)}\right)   \left(z-\frac{1}{2y}\right) ~,
\nonumber\\
\Omega_2^2 &=& \frac{y^2  (y-1)(y-x)}{x}   \left(
z-\frac{1}{2(y-x)}\right)  \left( z-\frac{1}{2(y-1)}\right) ~,
\nonumber\\
\Omega_3^2 &=& \frac{(y-1)^2  y  (y-x)}{(1-x)}   
 \left( z-\frac{1}{2y}\right)  \left( z-\frac{1}{2(y-x)}\right) ~,
\end{eqnarray}
together with an auxiliary variable
\begin{eqnarray}
z = \frac{x-2x  y + y^2 -2x  (1-x)  y'}{4y  (y-1)(y-x)} ~,
\end{eqnarray}
one can reduce
the first order equations (\ref{Omega})
to a single second order differential equation, i.e.
Painlev\'e VI equation :
\begin{eqnarray}
y''&=&
\frac{1}{2}\left(\frac{1}{y} + \frac{1}{y-1} + \frac{1}{y-x}\right){y'}^2
- \left(\frac{1}{x} + \frac{1}{x-1} + \frac{1}{y-x}\right)  y'
\nonumber\\&&
+ \frac{y (y-1)(y-x)}{x^2  (x-1)^2}  
 \left(
  \alpha + \beta \frac{x}{y^2} 
  + \gamma  \frac{x-1}{(y-1)^2} 
  + \delta  \frac{x  (x-1)}{(y-x)^2}
 \right) ~ ,\label{painleve}
\end{eqnarray}
with
$(\alpha,\beta,\gamma,\delta)=(1/8,\, -1/8,\, 1/8,\,3/8)$.

%%%%%%%%%%%%%%%%%%%%%%%%%%%%%%%%%%%%%%%%%%%%%%%%%%%%%%%%%%%%%%%%%%%%%
\sect{$G_2$-structure }
We assume the diagonal form of the Kaluza-Klein metric
(\ref{KK}),
\begin{eqnarray}
\bg_{diag}=dt^2+a^2(t) \sigma_1^2+b^2(t) \sigma_2^2+b^2(t) \sigma_3^2
+\alpha_1^2 (\phi^1)^2+\alpha_2^2 (\phi^2)^2+\alpha_3^2 (\phi^3)^2.
\end{eqnarray}
Provided the self-dual instanton $\phi^i=s_{ji}A^j+\tilde{\sigma}_i$,
the curvature $\Theta^i=d\phi^i+\frac{1}{2}\epsilon_{ijk}
\phi^j \wedge \phi^k$ is calculated as
\begin{eqnarray}
\Theta^i=-\frac{\Lambda}{3} s_{ji}\left(e^0 \wedge e^j+ \frac{1}{2}
\epsilon_{jk\ell} e^k \wedge e^{\ell} \right)~,\label{curvature}
\end{eqnarray}
where $(s_{ij}) \in $ SO(3) and $\{ e^{\mu};\, \mu=0,1,2,3 \}$
is the orthonormal basis of the Bianchi IX metric defined by
(\ref{orth}). We now introduce an orthonormal basis of the 
Kaluza-Klein metric : $\theta^i=\alpha_i \phi^i ~ (i=1,2,3)$
for the fiber metric, and $\theta^{\alpha} ~ (\alpha=4,5,6,7)$
are defined by the following equations,
\begin{eqnarray}
\Theta^1&=&\frac{\Lambda}{3}(\theta^4 \wedge \theta^5+\theta^6
\wedge \theta^7), \quad
\Theta^2=\frac{\Lambda}{3}(\theta^4 \wedge \theta^6+\theta^7
\wedge \theta^5), \nonumber\\
\Theta^3&=&\frac{\Lambda}{3}(\theta^4 \wedge \theta^7+\theta^5
\wedge \theta^6) 
\end{eqnarray}
and (\ref{curvature}). Then, the 3-form (\ref{G2}) can be written as
\begin{eqnarray}
\omega=\alpha_1 \alpha_2 \alpha_3 \phi^1 \wedge \phi^2 \wedge
\phi^3 +\frac{3}{\Lambda}
(\alpha_1 \phi^1 \wedge \Theta^1+\alpha_2 \phi^2 \wedge \Theta^2
+\alpha_3 \phi^3 \wedge \Theta^3)~.
\end{eqnarray}
Thus, the $G_2$-equation $d\omega=c * \omega$ reduces to the
algebraic equations ;
\begin{eqnarray}
\alpha_1&+&\alpha_2+\alpha_3=\frac{3 c}{2 \Lambda}\nonumber\\
\alpha_1 &\alpha_2& \alpha_3 +\frac{3}{\Lambda}
(-\alpha_1+\alpha_2+\alpha_3)=\frac{3 c}{\Lambda} \alpha_2 \alpha_3,
\end{eqnarray}
and the two equations obtained by cyclically permuting
$\alpha_1, \alpha_2, \alpha_3$. These reproduce the solution 
(\ref{sol}) and hence the metric (\ref{sol2}).

%%%%%%%%%%%%%%%%%%%%%%%%%%%%%%%%%%%%%%%%%%%%%%%%%%%%%%%%%%%%%%%%%%%%%%%%%%%

\end{document}